\documentclass[12pt]{article}%
\usepackage{pstricks}
\usepackage{color}
\usepackage{fullpage}
\usepackage{amsmath}%
\usepackage{amssymb}%
\usepackage{amscd}%
\usepackage{amsthm}
\usepackage{graphicx}
\usepackage{amssymb}
\usepackage{epstopdf}
\usepackage{cancel}
\usepackage{setspace}
\usepackage{slashed}
\usepackage[hang,small]{caption}

\def\half{\frac{1}{2}}

\newcommand{\be}{\begin{equation}}
\newcommand{\ee}{\end{equation}}

\newcommand{\mcal}[1]{\mathcal{#1}}

\newcommand{\ph}[1]{\phantom{#1}}

\newcommand{\duten}[3]{#1^{\ph{#2}#3}_{#2}}

\theoremstyle{definition}

\numberwithin{equation}{section}
\begin{document}

\begin{titlepage}
\bigskip
\rightline{}

\bigskip\bigskip\bigskip\bigskip
\centerline {\Large \bf { Complete Phase Diagrams for a Holographic  }}
\bigskip
\centerline{\Large \bf Superconductor/Insulator System}
\bigskip\bigskip
\bigskip\bigskip

\centerline{\large  Gary T. Horowitz and Benson Way}
\bigskip\bigskip
\centerline{\em Department of Physics, UCSB, Santa Barbara, CA 93106}
\centerline{\em  gary@physics.ucsb.edu, benson@physics.ucsb.edu}
\bigskip\bigskip
\begin{abstract}
The gravitational dual of an insulator/superconductor transition driven by increasing the chemical potential has recently been constructed. However, the system was studied in a probe limit  and only a part of the phase diagram was obtained. We include the backreaction and construct the complete phase diagram for this system. For fixed chemical potential there are typically two phase transitions as the temperature is lowered. Surprisingly, for a certain range of parameters, the system first becomes a superconductor and then becomes an insulator as the temperature approaches zero. As a byproduct of our analysis, we also construct the gravitational dual of a Bose-Einstein condensate of glueballs in a confining gauge theory. 
\end{abstract}
\end{titlepage}


\onehalfspacing
\begin{section}{Introduction}

The AdS/CFT correspondence \cite{Maldacena98,Gubser:1998bc,Witten:1998qj} has provided a framework where strongly coupled field theories can be studied using a weakly coupled gravity dual.  Recently, this correspondence has led to gravitational descriptions of superfluids and superconductors  \cite{Gubser08,HHH08a,HHH08b}. In the simplest realization,  a black hole in an Einstein-Maxwell-charged scalar theory becomes unstable to forming scalar hair below a critical temperature $T_c$.  In the dual field theory, this instability corresponds to a second order phase transition which  spontaneously breaks a $U(1)$ symmetry.  In this way, the field theory exhibits many properties shared by superconductors and superfluids.  This basic construction has been extended in various ways (for reviews see \cite{Hartnoll09,Herzog09,McGreevy:2009xe,Horowitz10}).   Such systems are called holographic superconductors.  
Above the phase transition, the black hole describes a conductor, so this is  a standard conductor/superconductor phase transition. 

In a recent paper \cite{NRT09}, it was shown that the same simple bulk theory with modified boundary conditions can describe an insulator/superconductor transition.  The new boundary conditions are simply that one direction is compactified to a circle (and fermions are antiperiodic around this circle). The insulator is described by the AdS soliton \cite{Witten98,Horowitz:1998ha}, which  has a mass gap and has previously been used to describe a confining vacuum of a dual gauge theory. In the well studied example of asymptotically $AdS_5 \times S^5$ solutions which are dual to ${\cal N} = 4 $ super Yang-Mills, the new boundary conditions give all fermions masses at tree level and the scalars pick up masses at one loop. The low energy theory is therefore a pure $2+1$-dimensional gauge theory which has a confining vacuum. The AdS soliton is the gravitational description of this state. For our purposes, 
 we will not consider any particular embedding in string theory and instead just view the soliton as describing an insulator.

It was shown in  \cite{NRT09} that if one adds a chemical potential $\mu$ to the AdS soliton, there is a second order phase transition at a critical value $\mu_c$ beyond which the charged scalar field turns on, even at zero temperature. It was also shown that for all $\mu \ge \mu_c$, the DC conductivity becomes infinite. So this is an insulator/superconductor transition.   

For small $\mu$, it is known that there is another phase transition when one increases the temperature \cite{Surya:2001vj}. This is a first order phase transition between the AdS soliton and the AdS black hole\footnote[1]{This phase transition is the planar analogue of the Hawking-Page transition for global AdS \cite{HP83}.}.  In  gauge theory language, this is a confining/deconfining phase transition \cite{Witten98}, while in the condensed matter interpretation it is a insulator/conductor transition.  Therefore, there are a total of four phases in this system given by the AdS soliton, the AdS black hole, and their superconducting phases.  Thus, just by compactifying one direction, the simple holographic superconductor model gets a much richer  phase diagram.

The analysis in \cite{NRT09} was done in the probe limit where the backreaction of the matter fields is ignored.  This is justified in the limit of large charge $q$ on the scalar field with $q\mu$ fixed. The transition in which the soliton develops a nonzero  scalar field was studied in this limit, but since the probe limit requires $\mu \ll 1$, the full phase diagram was not constructed. 

In this paper, we extend this previous work and include the backreaction of the matter fields on the soliton geometry and black hole. We also compute the full phase diagrams as a function of temperature and chemical potential for various $q$. These are given in Figure 6.  For all $q >1$, we find that the black hole phases never dominate at low temperature. The system always prefers either the soliton insulator or soliton superconductor as $T \rightarrow 0 $. The four phases  typically meet in two triple points where three of the phases can coexist. As we lower $q$, the two triple points merge into one ``quadruple" point and then separate, passing through each other. This produces a new phase boundary in which 
 the system goes from a conductor to a superconductor and then to an insulator as the temperature is lowered!  This is very different from the superconductor/insulator transition studied in \cite{NRT09}. In the previous case, the transition was second order and occurred by decreasing the chemical potential at fixed temperature. The new superconductor/insulator transition is first order and occurs when decreasing the temperature at fixed chemical potential.

As an aside, we also consider the limiting case of a neutral scalar field. When $q=0$, the AdS soliton is never unstable to turning on a static scalar field. However, there are finite energy solutions in which the geometry is static and the scalar field has pure harmonic time dependence. These are planar analogues of boson stars.  (For a review of boson stars see \cite{SM03}, and for a discussion of spherical  boson stars in asymptotically AdS spacetimes see \cite{AR03}.)  Like typical boson star solutions, we find that there is a maximum mass that can be supported.  In terms of a dual gauge theory, these solutions correspond to Bose-Einstein condensates of glueballs.

The paper is organized as follows.  In the following section, we write down the equations of motion for our systems and discuss the solutions with no  scalar field.  In section three, we numerically construct the soliton solutions with scalar backreaction, including the $q=0$ case.  We then construct the black hole solutions with scalar hair in section four.  In section five, we assemble the phase diagrams of our system.  The last section summarizes our results.

\end{section}
\begin{section}{The Bulk Equations and Solutions without Scalar}

We begin with a five-dimensional Einstein-Maxwell-scalar action:
\begin{equation}\label{action}
S=\int d^5x\sqrt{-g}\left(R+\frac{12}{L^2}-\frac{1}{4}F^{\mu\nu}F_{\mu\nu}-|D\psi|^2 -m^2|\psi|^2\right)\;.
\end{equation}
Here, we are writing $F=dA$ and $D_\mu=\nabla_\mu-iqA_\mu$.  The equations of motion obtained from this action are the scalar equation
\begin{equation}
-D_\mu D^\mu\psi+m^2\psi=0\;,
\end{equation}
Maxwell's equations
\begin{equation}
\nabla^\mu F_{\mu\nu}=iq[\psi^*D_\nu\psi-\psi D_\nu^*\psi^*]\;,
\end{equation}
and Einstein's equations
\begin{align}
R_{\mu\nu}-\half g_{\mu\nu}R-\frac{6g_{\mu\nu}}{L^2}&=\half F_{\mu\lambda}\duten{F}{\nu}{\lambda}+\half \left(D_\mu\psi D_\nu^*\psi^*+D_\nu\psi D_\mu^*\psi^*\right)\notag\\
&\qquad\qquad-\frac{g_{\mu\nu}}{2}\left(\frac{1}{4}F^{\rho\sigma}F_{\rho\sigma}+m^2|\psi|^2+|D\psi|^2\right)\;.
\end{align}
First, let's consider solutions with $\psi=0$ and $A_t=\phi(r)$.  One solution with planar symmetry is the AdS Reissner-N\"{o}rdstrom black hole, given by \cite{Hartnoll09}
\begin{equation}\label{RNmetric}
ds^2=-f(r)dt^2+\frac{L^2dr^2}{f(r)}+r^2(dx^2+dy^2+dz^2)\;,
\end{equation}
where
\begin{equation}\label{RNf}
f(r)=r^2\left[1-\left(1+\frac{\mu^2}{3r_+^2}\right)\left(\frac{r_+}{r}\right)^4+\frac{\mu^2}{3r_+^2}\left(\frac{r_+}{r}\right)^6\right]\;,
\end{equation}
and
\begin{equation}
\phi(r)=\mu\left[1-\left(\frac{r_+}{r}\right)^2\right]\;.
\end{equation}
We have written the metric in this form so that $r_+$ is the horizon radius and $\mu$ is the chemical potential of the dual field theory.  The temperature of this black hole is
\begin{equation}\label{bhtemp}
T=\frac{r_+}{\pi L}\left(1-\frac{\mu^2}{6r_+^2}\right)\;.
\end{equation}
We will later want to make $z$ a periodic coordinate.  Any period can be chosen for $z$.  

There is another planar solution, the AdS soliton, which is given by
\begin{equation}
ds^2=\frac{L^2dr^2}{f(r)}+r^2(dx^2+dy^2-dt^2)+f(r)d\eta^2,\qquad f(r)=r^2-\frac{r_0^4}{r^2}\;.
\end{equation}
and $\phi = 0$.  This solution can be obtained by a double Wick rotation of the AdS Schwarzschild black hole. That is, we set $\mu=0$ in \eqref{RNmetric} and \eqref{RNf} and make the substitutions $t\rightarrow i\eta$ and $z\rightarrow it$.  The geometry resembles a cigar with the tip at $r=r_0$.  Unlike the black hole, in order to avoid a conical singularity at the tip, $\eta$ must be chosen to have a period of 
\be
\gamma=\frac{\pi L}{r_0}\;.
\ee
  The soliton can be considered at any temperature.  That is, we are free to give imaginary time any period $\beta=1/T$ we wish.  

Due to the periodicity of $\eta$, the theory dual to the AdS soliton lives on a space with one direction compactified to a circle with length $\gamma$. Because the AdS soliton only exists for $r\ge  r_0$, this dual field theory is confining and has a mass gap.  Consider this theory at finite temperature. There are two phases given by the Euclidean black hole (with $\mu=0$ and $z$ periodic with period $\gamma$) and the Euclidean soliton.  These are actually the same geometry, differing only in the interpretation of which $S^1$ at infinity is Euclidean time and which is a spacelike direction. For the black hole, the Euclidean time direction is contractible, while for the soliton, the spatial direction is contractible.  Since the free energy is computed from the Euclidean action, there is a phase transition when the two circles have the same length: $\beta = \gamma$. This is a first order transition, exactly analogous
to the Hawking-Page transition for global AdS \cite{HP83}. 

One can extend this to nonzero chemical potential by noticing that one can add $\phi = \mu$ to the soliton solution without changing the metric. As we increase $\mu$, there is an instability which turns on the charged scalar field (in both the soliton and black hole phases). This was analyzed in the probe limit ($q\rightarrow\infty$, keeping $q\psi$ and $q\phi$ fixed) in which backreaction on the metric is  ignored in \cite{NRT09}.  We would like to move beyond the probe limit and study the phase transitions with full backreaction.  To do this, we must solve the full set of differential equations, which we turn to in the following sections.  
\end{section}

\begin{section}{AdS Soliton with Scalar}
\begin{subsection}{Equations of Motion and Boundary Conditions}
Now let us now consider the  soliton solutions with $\psi\neq0$.  We will work in units in which $L=1$.  We are interested in including backreaction so we choose the metric ansatz
\begin{equation}
ds^2 =r^2\left(e^{A(r)}B(r)d\eta^2+dx^2+dy^2-e^{C(r)}dt^2\right)+\frac{dr^2}{r^2B(r)}\;.
\end{equation}
We require that $B$ vanish at some radius $r_0$, which is the tip of the soliton.
As in the case without a scalar field, smoothness at the tip requires that $\eta$ be periodic with period
\begin{equation}
\gamma=\frac{4\pi e^{-A(r_0)/2}}{r_0^2B'(r_0)}\;.
\end{equation}
We would like to consider solutions of the form
\begin{equation}\label{Apsiansatz}
A_t=\phi(r),\qquad \psi=\psi(r)e^{-i\omega t}\;.
\end{equation}
For $q\neq0$, there is a symmetry
\begin{equation}
\psi\rightarrow\psi e^{iat},\qquad \phi\rightarrow\phi+\frac{a}{q},
\end{equation}
that lets us set $\omega=0$.  If $q=0$, then $\phi$ decouples from $\psi$ and we cannot gauge away $\omega$.  

With this ansatz, the scalar and Maxwell equations become
\begin{equation}
\psi''+\left(\frac{5}{r}+\frac{A'}{2}+\frac{B'}{B}+\frac{C'}{2}\right)\psi'+\frac{1}{r^2B}\left(\frac{e^{-C}(\omega+q\phi)^2}{r^2}-m^2\right)\psi=0\;,
\end{equation}
\begin{equation}
\phi''+\left(\frac{3}{r}+\frac{A'}{2}+\frac{B'}{B}-\frac{C'}{2}\right)\phi'-\frac{2\psi^2q(\omega+q\phi)}{r^2B}=0\;.
\end{equation}

For simplicity, we choose $m^2=-\frac{15}{4}$.  Our analysis can be generalized to any mass above the Breitenlohner-Freedman bound \cite{BF82} $m^2>-\frac{(D-1)^2}{4}=-4$.  Near the boundary $r=\infty$, the scalar and Maxwell equations have the form
\begin{equation}\label{psibndry}
\psi=\frac{\psi^{(1)}}{r^{3/2}}+\frac{\psi^{(2)}}{r^{5/2}}+\ldots\;,
\end{equation}
\begin{equation}\label{phibndry}
\phi=\mu-\frac{\rho}{2r^2}+\ldots\;.
\end{equation}
The quantities $\mu$ and $\rho$ are interpreted as the chemical potential and the charge density in the dual field theory, respectively.  At our value of $m^2$, the $\psi^{(1)}$ and $\psi^{(2)}$ terms are both normalizable so we have a choice of boundary conditions.  For numerical simplicity, we only consider the case $\psi^{(1)}=0$.  The constants $\psi^{(1)}$ and $\psi^{(2)}$ can be used to define operators on the dual field theory with mass dimension $\Delta=3/2$ and $\Delta=5/2$, respectively.  Up to a normalization, the expectation value of the scalar operator $\mcal O$ is given by
\begin{equation}\label{operator}
\langle\mcal O\rangle=\psi^{(2)}\;.
\end{equation}

The nontrivial components of Einstein's equations are the $tt$, $rr$, $\eta\eta$, and $xx$.  These equations are not all independent.  The $xx$ component, for example, can be derived from combinations of the other equation of motion and their derivatives.  Therefore, we only need three linearly independent combinations of these equations.  Subtracting the $\eta\eta$ equation from the $rr$ equation allows us to solve for $A'$:
\begin{equation}\label{Aeq}
A'=\frac{2r^2C''+r^2C'^2+4rC'+4r^2\psi'^2-2e^{-C}\phi'^2}{r(6+rC')}\;.
\end{equation}
The remaining two equations come from the $xx-tt$ component
\begin{equation}
C''+\half C'^2+\left(\frac{5}{r}+\frac{A'}{2}+\frac{B'}{B}\right)C'-\left(\phi'^2+\frac{2(\omega+q\phi)^2\psi^2}{r^2B}\right)\frac{e^{-C}}{r^2}=0
\end{equation}
and $\eta\eta+tt-xx$
\begin{equation}
B'\left(\frac{3}{r}-\frac{C'}{2}\right)+B\left(\psi'^2-\half A'C'+\frac{e^{-C}\phi'^2}{2r^2}+\frac{12}{r^2}\right)+\frac{1}{r^2}\left(\frac{e^{-C}(\omega+q\phi)^2\psi^2}{r^2}+m^2\psi^2-12\right)=0\;.
\end{equation}
Note that we can substitute \eqref{Aeq} into the other equations of motion to eliminate $A$.  This does not change the order of our differential equations. We can then integrate \eqref{Aeq} separately after solving the other equations.   

 By considering a series solution about the tip of the soliton, $r = r_0$, and using $B(r_0) = 0$, we are left with four independent parameters: $r_0$, $\psi(r_0)$, $\phi(r_0)$, $C(r_0)$.  However, there are two useful symmetries.
The scaling symmetry
\begin{equation}\label{solitonscaling}
r\rightarrow ar,\qquad (\eta,x,y,t)\rightarrow(\eta,x,y,t)/a,\qquad\phi\rightarrow a\phi
\end{equation}
can be used to set $r_0=1$ for performing numerics.  In general, different solutions obtained this way will have different periods $\gamma$ for the $\eta$ coordinate.  In order to make appropriate comparisons between different solutions,  the boundary geometry must be the same.  Therefore, we will later use \eqref{solitonscaling} again to set all of the periods $\gamma$ equal.  
Similarly, the symmetry
\begin{equation}
e^{-C}\rightarrow a^2e^{-C},\qquad t\rightarrow at,\qquad\phi\rightarrow \phi/a
\end{equation}
lets us pick an arbitrary value of $C(r_0)$.  After solving the differential equations, we can  read off the value of $C(\infty)$ and use this symmetry to satisfy the asymptotic condition $C(\infty)=0$.  

 If $q\neq0$, we can choose $\phi(r_0)$ as a shooting parameter, which we tune to satisfy the remaining asymptotic condition, $\psi^{(1)}=0$.  If $q=0$, $\phi$ decouples from $\psi$ (and will be set to zero). However precisely in this case, $\omega$ cannot be gauged away and becomes our shooting parameter.  

So for each choice of $q$ and $\psi(r_0)$, we can solve the equations of motion for $\psi$, $\phi$, $B$, and $C$.  The remaining function $A$ can be found by integrating \eqref{Aeq} with the condition $A(\infty)=0$.  Thus for fixed $q$, we have a one-parameter family of solutions.
After solving the equations, the AdS/CFT correspondence allows us to obtain the chemical potential $\mu$ and the charge density $\rho$ from the asymptotic behaviour of $\phi$ according to \eqref{phibndry}.  We can also obtain $\langle\mcal O\rangle$ through the asymptotic behaviour of $\psi$ through \eqref{psibndry} and \eqref{operator}.  
\end{subsection}

\begin{subsection}{$q\neq0$}

Our main interest is the case when the scalar field is charged: $q\neq0$.  One finds that when $\mu$ is small enough, there are no solutions for $\psi$ which satisfy our boundary condition. However when one increases $\mu$, such solutions exist. It is convenient to express the results in terms of scale-invariant quantities. The condensate $\langle\mcal O\rangle$ has dimension $5/2$, the chemical potential has dimension one, and the charge density has dimension three. Using the length of the circle at infinity, $\gamma$, as a scale  we plot  $\langle\mcal O\rangle^{2/5}\gamma$ and $\rho\gamma^3$ as functions of $\mu\gamma$ in    Figure \ref{solitonq2}  for $q=2$.  We have found that the plots for $q=10$, $5$, and $1.2$ are qualitatively similar. 
\begin{figure}[h]
  \centering
  \includegraphics[scale=.85]{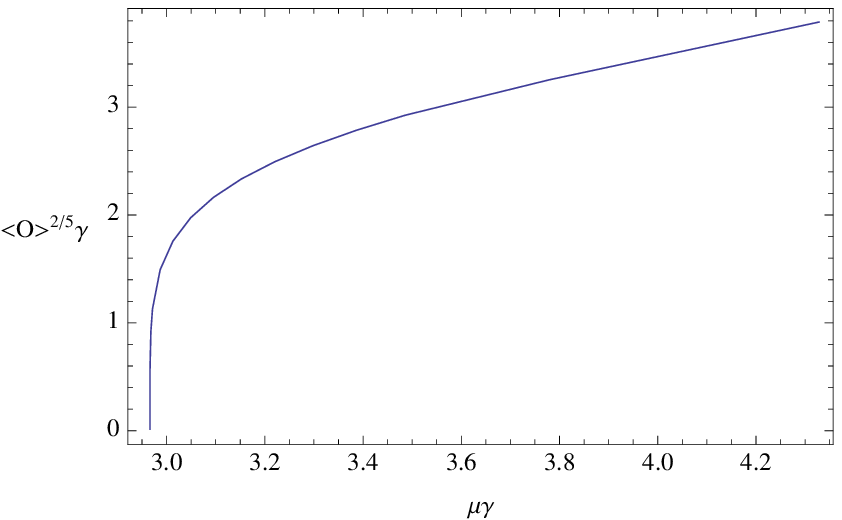}$\qquad\qquad$
  \includegraphics[scale=.85]{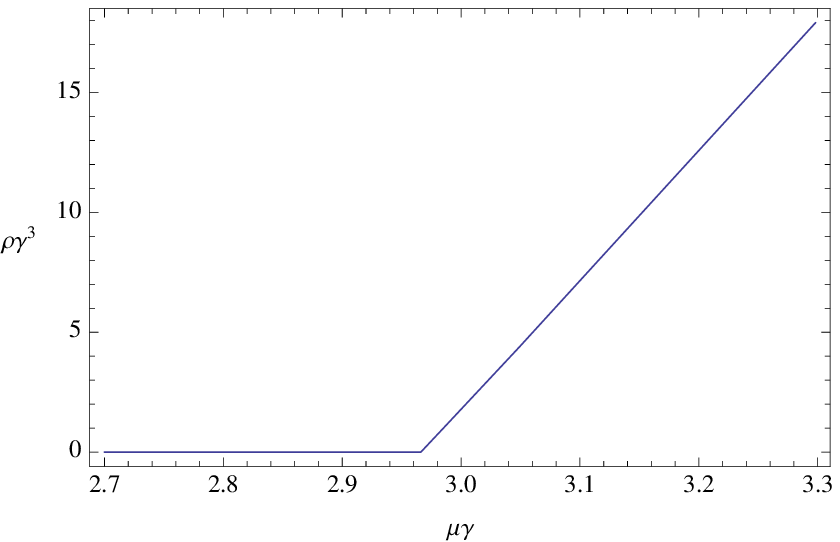}
  \caption{The value of the condensate (left), and the charge density (right) as a function of chemical potential, for the soliton with $q=2$.}
  \label{solitonq2}
\end{figure}

The main conclusion of Figure 1 is that a phase transition occurs at a critical value of $\mu$.  The transition is second-order, which can be seen from the $\rho$-$\mu$ plot.  If one computes the optical conductivity in the AdS soliton background (with no scalar) one finds that ${\rm Re}[\sigma(\omega)]$ is a series of delta functions at nonzero frequency \cite{NRT09}.  In particular, the DC conductivity is zero, so this describes an insulator. When the scalar field is nonzero, there is an additional delta function at $\omega = 0$  indicating the onset of superconductivity \cite{NRT09}. Thus, we have a holographic superconductor/insulator phase transition that can exist at strictly zero temperature.  

There is an additional complication when $q\lesssim1.2$.  Recall that our one parameter family of solutions can be labelled by the value of the scalar field at the tip of the soliton. For $q\gtrsim1.2$, as one increases $\psi(r_0)$, $\mu\gamma$ monotonically increases. So every value of the chemical potential has at most one soliton solution with scalar. When $q\lesssim1.2$, $\mu\gamma$ does not monotonically increase with $\psi(r_0)$ and there are up to three different solutions for the same chemical potential. The free energy\footnote{See section 5 for more discussion on the computation of free energy.} versus chemical potential plots develop ``swallow tails" (see Figure \ref{solitonfq1}) common in  first order phase transitions.  A similar effect has been observed in spherical Reissner-N\"ordstrom AdS black holes (when the total charge is fixed rather than the chemical potential)
  \cite{CEJM}.  This new phase transition first appears inside the superconducting phase  at some $\mu_{dis}$.  The charge density $\rho$ and condensate $\langle\mcal O\rangle$ are also discontinuous at $\mu_{dis}$ (see Figure \ref{solitonq1}). It is not clear what physically distinguishes these two phases of the superconductor. As one lowers $q$, the first order phase transition moves over and  coincides with the superconductor/insulator transition when $q \approx 1$.

\begin{figure}[h]
  \centering
   \includegraphics[width=3in]{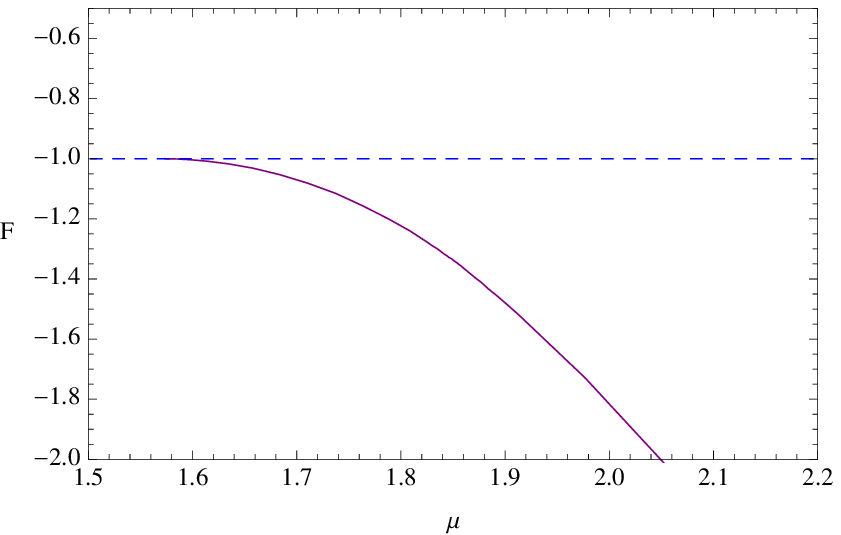}\qquad
  \includegraphics[width=3in]{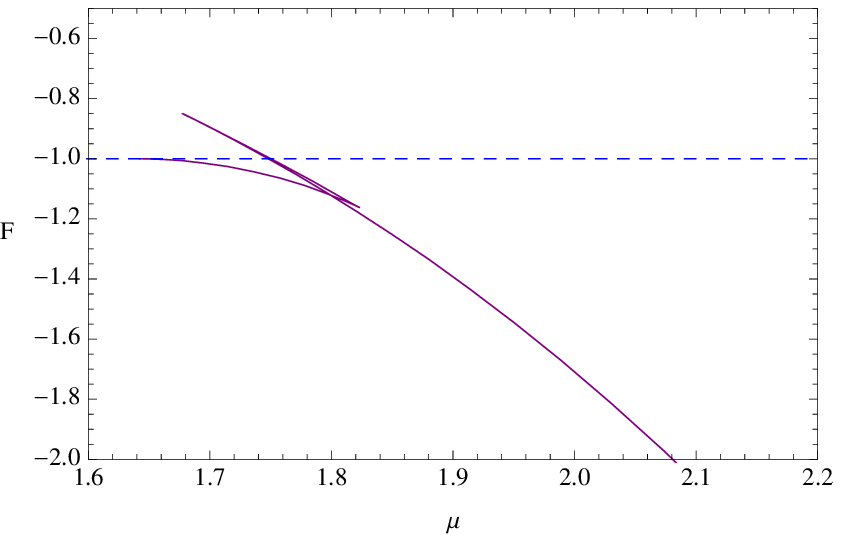}\qquad
  \linebreak
  \linebreak
  \includegraphics[width=3in]{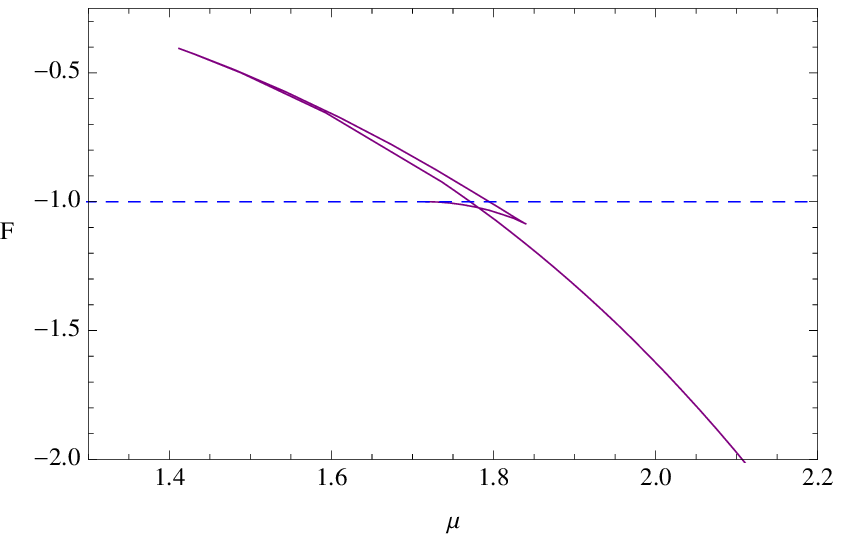}\qquad
  \includegraphics[width=3in]{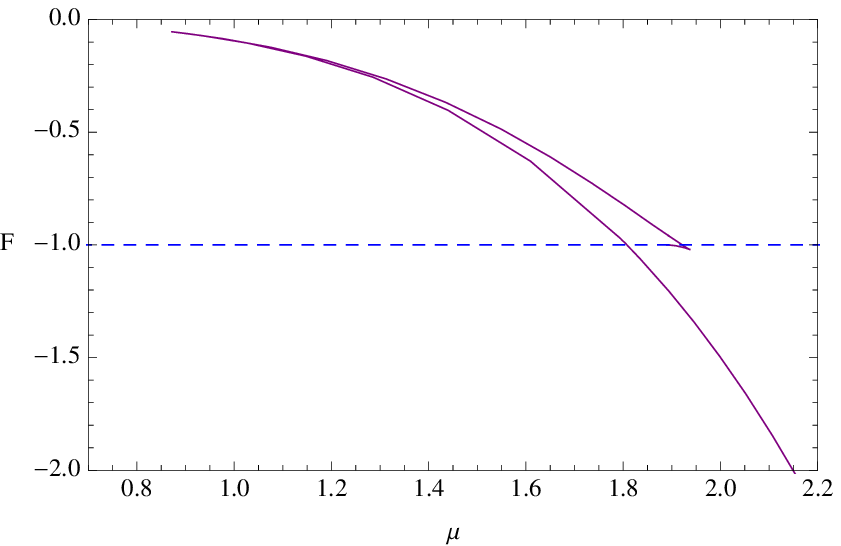}
  \caption{The free energy densities of the soliton with scalar (solid purple), and the soliton without scalar (dashed blue), against chemical potential with $q=1.2$ (top left), $q=1.15$ (top right), $q=1.1$ (bottom left), and $q=1$ (bottom right).  Here, we have scaled $\gamma=\pi$.   The $q=1.2$ plot shows the typical secord-order phase transition seen in the probe limit.  The $q=1.15$ and $q=1.1$ plots have second order phase transitions, but a discontinuity within the superconducting phase.  The $q=1$ plot shows a first order phase transition.}
  \label{solitonfq1}
\end{figure}
\begin{figure}[h]
  \centering
     \includegraphics[scale=.9]{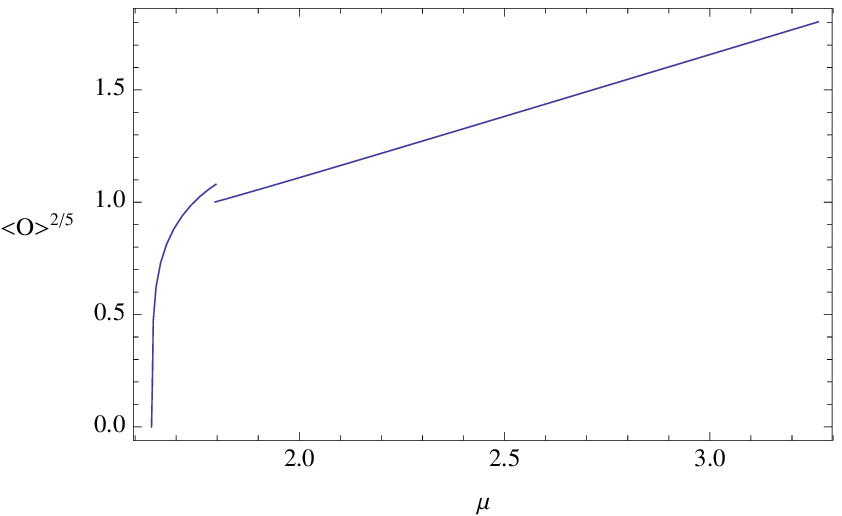}\qquad
  \includegraphics[scale=.82]{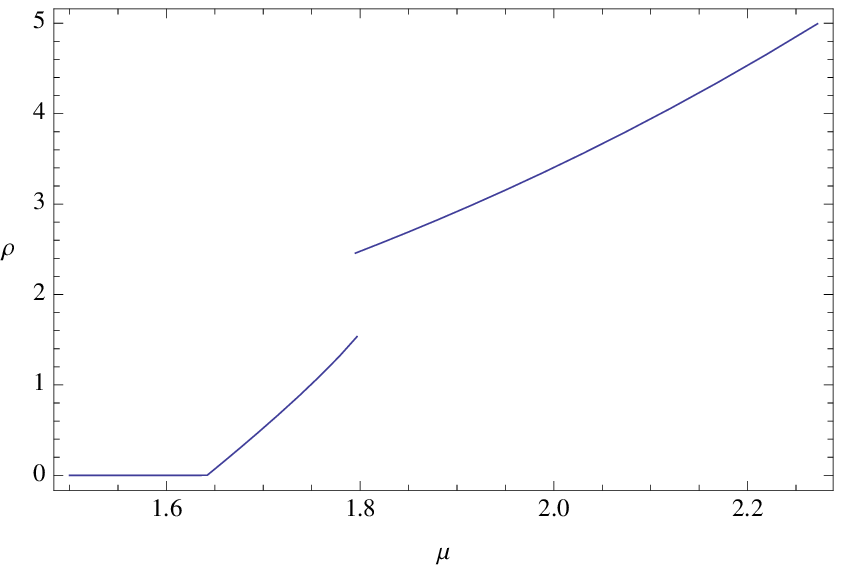}\qquad
  \linebreak
  \linebreak
  \includegraphics[scale=.9]{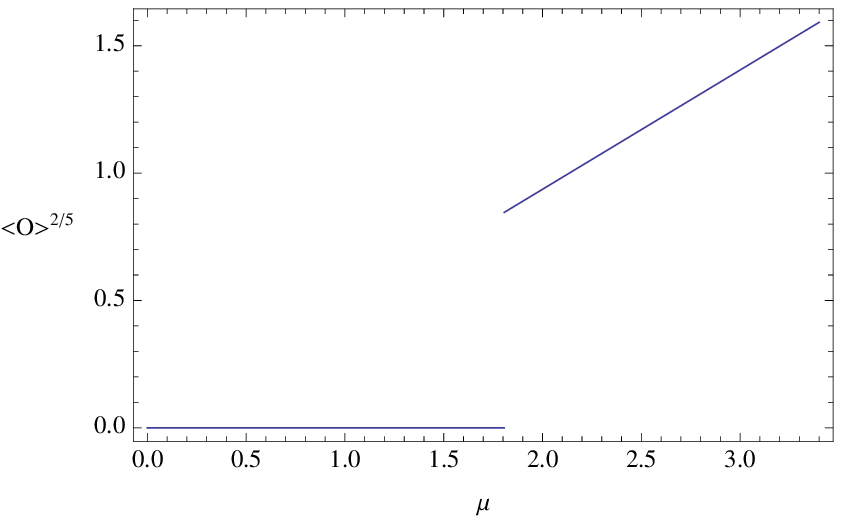}\qquad
  \includegraphics[scale=.8]{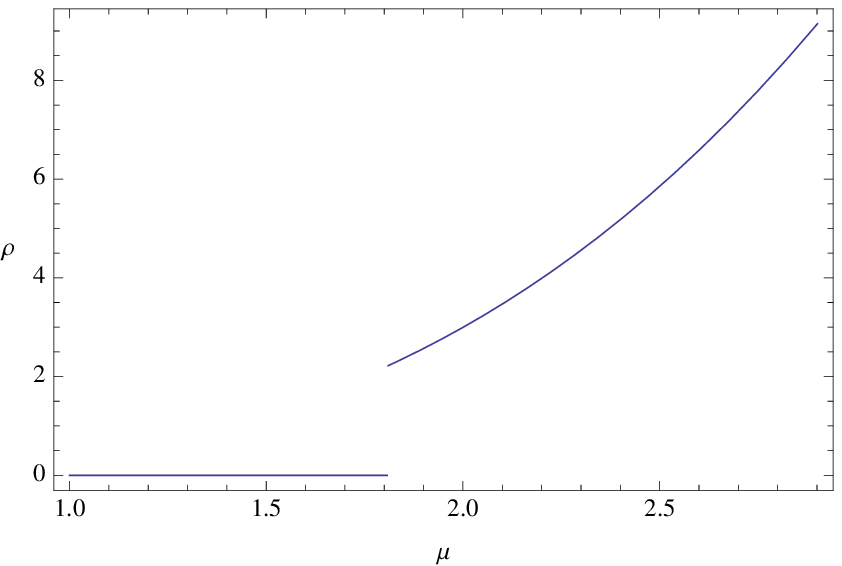}
  \caption{The value of the condensate (left plots), and the charge density (right plots) as a function of chemical potential, for the AdS soliton with $q=1.15$ (top plots) and $q=1$ (bottom plots).  Here, we have scaled $\gamma=\pi$.  Note that the $q=1.15$ plots have a second order phase transition into the superconducting phase (at $\mu\approx1.71$), but within the superconducting phase (at $\mu\approx 1.8$), there's a discontinuity in the charged density and condensate.   The $q=1$ plots give a first order phase transition. }
  \label{solitonq1}
\end{figure}
\end{subsection}
\begin{subsection}{$q=0$}
Now let's consider $q=0$ and turn off the chemical potential: $\mu = 0$.  The Maxwell field  drops out of the problem and we are left with just gravity coupled to a complex scalar field. We look for solutions of  the form $\psi=\psi(r)e^{-i\omega t}$ with $\omega\neq0$.  For every choice of $\psi(r_0)$, one can adjust $\omega$ so that $\psi(r)$ is normalizable at infinity. Therefore, our solutions are boson stars in the sense that we have a static spacetime solution with a complex scalar varying harmonically with time.  Our solutions, however, have planar symmetry while most boson stars that have been studied have spherical symmetry.

The energy of the boson stars are  plotted in Figure~\ref{solitonq0}.  The AdS soliton itself has negative energy, i.e., energy below $AdS_5$.  As we increase $\psi(r_0)$ the energy grows to a positive maximum and then decreases. This is similar to previous studies of 
 boson star solutions, in  that there is a maximum mass that can be supported.  We  find that the condensate also has a maximum value.  These results are shown in Figure~\ref{solitonq0}.  

In typical boson stars, $\psi_c$, the value of $\psi(r_0)$ corresponding to the maximum mass, is a boundary between stable and unstable solutions.   Solutions with $\psi(r_0)<\psi_c$ are stable while solutions with $\psi(r_0)>\psi_c$ are unstable \cite{AR03}.  The unstable solutions will evolve into a Schwarzschild black hole.  We believe that our solutions will behave similarly, although we do not analyze stability here.   One simple check is that the energy in the unstable branch is always positive and hence there is a black hole with  the same energy for it to evolve into.

 When $\psi(r_0)$ is small, the backreaction is negligible and one is solving the linearized scalar equations on the AdS soliton background. In terms of the dual gauge theory, this describes a spin 0 glueball state above the confining vacuum. By increasing $\psi(r_0)$ and including backreaction, one is describing the gravitational dual of a Bose-Einstein condensate of glueballs. Figure 4 shows that these condensates remain coherent and do not thermalize even when their energy is nearly double  that required to drive a deconfinement transition. 
 
Since the double analytic continuation of the AdS soliton is a planar black hole, it is natural to ask what happens if one does the same analytic continuation on the boson star. Setting $t \rightarrow iz$ and $\eta \rightarrow it$, the scalar becomes $\psi(r, z)=\psi(r)e^{\omega z}$ and the metric becomes a translationally invariant black hole. It thus appears that one can have inhomogeneous scalar hair outside a homogeneous black hole! To understand this strange solution, we must take a closer look at the behavior of the Lagrangian under analytic continuation. The scalar field in the boson star is complex, so the analytic continuation results in two real scalar fields. Starting with $\psi_1 = \psi(r) \cos{\omega t}$ and $\psi_2 = \psi(r) \sin{\omega t}$ we see that $\psi_1$ is the expected exponentially growing solution outside the horizon, but $\psi_2 = i\psi(r) \sinh{\omega z}$ is pure imaginary. This means that it has negative kinetic energy and the theory does not have a stable ground state.

\begin{figure}[h]
  \centering
  \includegraphics[scale=.85]{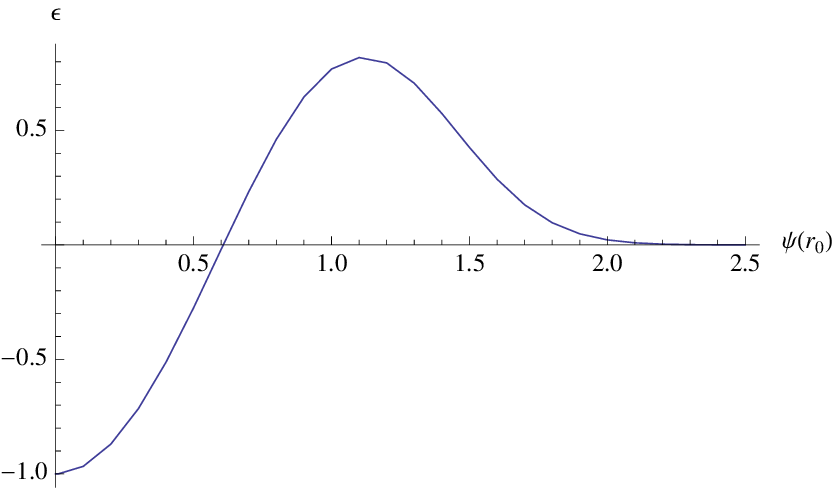}$\qquad\qquad$
  \includegraphics[scale=.85]{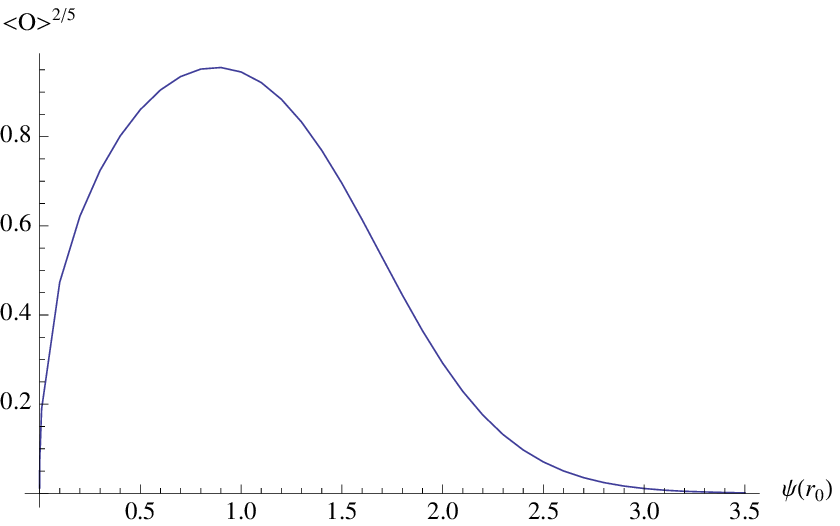}
  \caption{The energy density (left), and the condensate (right) as a function of scalar field at the tip, for the AdS soliton with neutral scalar field excited.  Here, we have scaled the period of the $\eta$ coordinate to be $\gamma=\pi$.  These solutions have a maximum mass and condensate.}
  \label{solitonq0}
\end{figure}
\end{subsection}
\end{section}

\begin{section}{AdS Charged Black Hole with Scalar}

In this section we construct asymptotically AdS charged black holes with scalar hair in five dimensions with $m^2 = -15/4$ and various values of $q$. It is convenient to use the metric ansatz
\begin{equation}
ds^2=-g(r)e^{-\chi(r)}dt^2+\frac{dr^2}{g(r)}+r^2(dx^2+dy^2+dz^2)\;,
\end{equation}
which has a temperature
\begin{equation}\label{bhsctemp}
T=\frac{g'(r_+)e^{-\chi(r_+)/2}}{4\pi}\
\end{equation}
and entropy
\begin{equation}\label{bhscs}
S=4\pi r_+^3V_3=4\pi r_+^3\gamma V_2\;
\end{equation}
where the horizon, $r_+$, is defined by $g(r_+)=0$.  
To match with the soliton, we will periodically identify the $z$ coordinate with period $\gamma=\pi$.  Then with the same ansatz for $A_\mu$ and $\psi$ as the soliton \eqref{Apsiansatz}, and setting  $\omega=0$, the scalar and Maxwell's equations are
\begin{equation}
\psi''+\left(\frac{3}{r}-\frac{\chi'}{2}+\frac{g'}{g}\right)\psi'+\frac{1}{g}\left(\frac{e^\chi q^2\phi^2}{g}-m^2\right)\psi=0\;,
\end{equation}
\begin{equation}
\phi''+\left(\frac{3}{r}+\frac{\chi'}{2}\right)\phi'-\frac{2q^2\psi^2}{g}\phi=0\;,
\end{equation}
while the $tt$ and $tt-rr$ components of Einstein's equations become
\begin{equation}
\chi'+\frac{2r\psi'^2}{3}+\frac{2re^\chi q^2\phi^2\psi^2}{3g^2}=0\;,
\end{equation}
\begin{equation}
g'+\left(\frac{2}{r}-\frac{\chi'}{2}\right)g+\frac{re^\chi \phi'^2}{6}+\frac{m^2r\psi^2}{3}-4r=0\;.
\end{equation}
The $xx$ component of Einstein's equation can be derived by differentiating the other equations of motion.  As before, we choose $m^2=-\frac{15}{4}$.  Note that \eqref{psibndry} and \eqref{phibndry} still hold in this case.  

Now we consider boundary conditions.  In addition to $g(r_+) = 0$, we must impose $\phi(r_+)=0$ in order for $g^{\mu\nu}A_\mu A_\nu$ to remain finite at the horizon.  Then the independent parameters at the horizon are $r_+$, $\psi(r_+)$, $\phi'(r_+)$, and $\chi(r_+)$.  As in the case for the soliton, we can use the scaling symmetries
\begin{equation}\label{bhscaling}
r\rightarrow ar,\qquad(\tau,x,y,t)\rightarrow(\tau,x,y,t)/a,\qquad g\rightarrow a^2g,\qquad \phi\rightarrow a\phi\;,
\end{equation}
and 
\begin{equation}
e^\chi\rightarrow a^2e^\chi,\qquad t\rightarrow at,\qquad \phi\rightarrow \phi/a
\end{equation}
to set $r_+=1$ and $\chi(\infty)=0$.  We will later use \eqref{bhscaling} again to rescale $\mu$ in order to make comparisons with the soliton solutions.  We choose $\phi'(r_+)$ as a shooting parameter to set $\psi^{(1)}=0$.  

After solving the differential equations numerically, we can plot the scale-invariant quantities $\langle \mcal O\rangle^{2/5}/\mu$ and $\rho/\mu^3$ versus $T/\mu$, as shown in Figure \ref{bhq2}.  These scale invariant quantities are equivalent to those scaled to $\mu=1$.  These results are similar to results for the four-dimensional superconductors studied in \cite{HHH08a,HHH08b}.  The scalar turns on quickly below some critical temperature through a second-order phase transition.  Unlike the AdS soliton, the phase transition remains second order  even for $q<1$. (We have checked as low as $q=0.5$).  
\begin{figure}[h]
  \centering
  \includegraphics[scale=0.85]{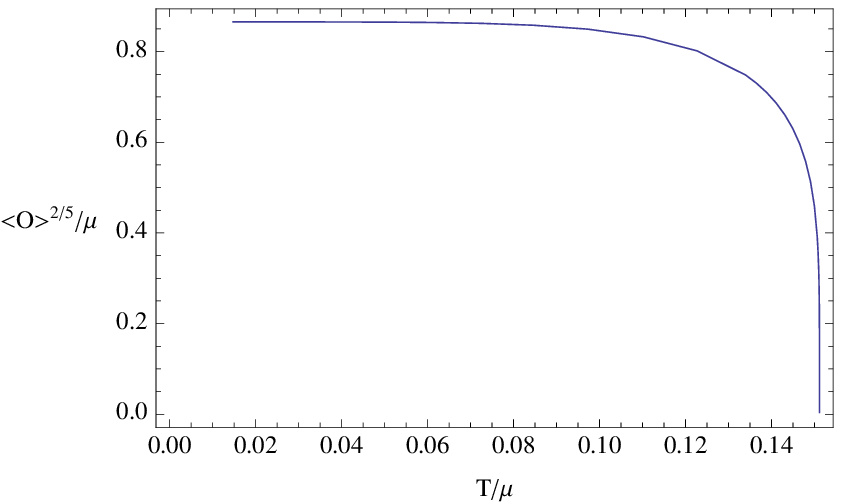}\qquad\qquad
  \includegraphics[scale=0.85]{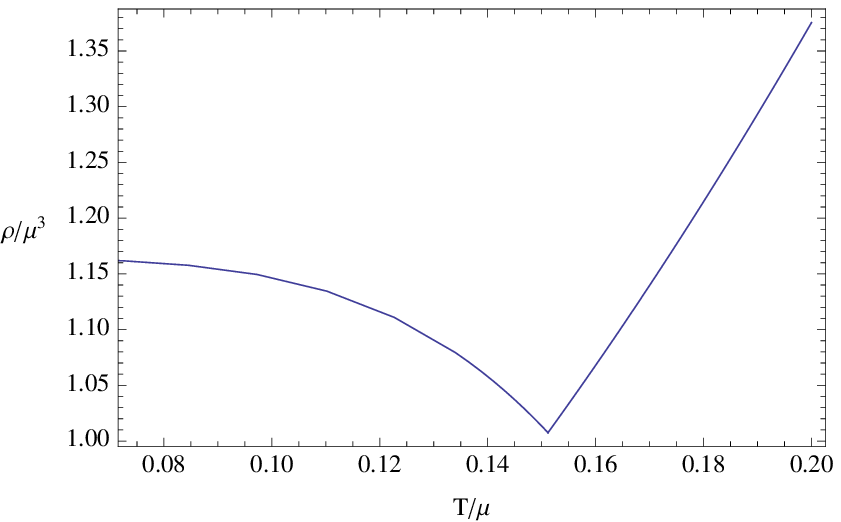}
  \caption{The value of the condensate (left), and the charge density (right) as a function of chemical potential, for the AdS black hole with $q=2$.}
  \label{bhq2}
\end{figure}

\end{section}

\begin{section}{Phase Diagrams}

We have seen that the system dual to our bulk Einstein-Maxwell-scalar theory has four phases. There is an insulating phase described by the AdS soliton, a conducting phase described by the charged black hole, a superconducting phase described by the hairy black hole, and a second superconducting phase described by the soliton with nonzero scalar field. With slight abuse of terminology,  we will refer to the last two phases as the black hole superconductor and soliton superconductor respectively. A key difference between these two superconducting phases is that for the black hole superconductor, the low temperature optical conductivity has a gap at low frequency but approaches the normal state (nonzero) conductivity at larger frequency. For the soliton superconductor, the real part of the conductivity is just a series of delta functions.

In this section, we construct the complete phase diagram as a function of temperature and chemical potential. There is a third scale in the problem, the size of the compact direction at infinity, which we fix to be $\gamma = \pi$. This removes the usual scaling symmetry. To construct the phase diagram, we must compute the Gibbs free energy of each system in the grand canonical ensemble.  This can be obtained from the classical gravity action \eqref{action} of the Euclidean solutions: 
\begin{equation}
I_G=\beta \Omega=\beta(E-TS-\mu Q)\;,
\end{equation}
where $E$ is the total energy.  Written in terms of densities,
\begin{equation}\label{fsoliton}
F=\frac{\Omega}{\gamma V_2}=\epsilon-Ts-\mu\rho\;.
\end{equation}

Let's first consider the AdS soliton without the scalar field. The entropy and charge density vanish for this solution so the free energy is given (in our units) by\footnote{We have dropped the $1/16\pi G$ factor in front of the action and set $L=1$. Keeping these factors, all of the free energies are proportional to $L^3/G$ which is $O(N^2)$ in the standard mapping to super Yang-Mills theory. This is different from the Hawking-Page transition in global AdS where the free energy is $O(1)$ at low temperatures.}
\begin{equation}
F_{sl}=\epsilon_{sl}=-\frac{\pi^4}{\gamma^4}=-1\;,
\end{equation}
where in the last equality we have used the fact that $\gamma=\pi$.  
The free energy of the AdS soliton with a charge scalar is given by
\begin{equation}\label{fsolitonsc}
F_{slsc}=\epsilon_{slsc}-\mu\rho_{slsc}\;.
\end{equation}
The free energy of these solutions is independent of temperature, so a phase boundary between them appears along a line of constant chemical potential.

Now consider the Reissner-N\"ordstrom AdS black hole.  We can calculate the free energy through \cite{Hartnoll09}
\begin{equation}\label{fbh}
F_{bh}=-r_+^4\left(1+\frac{\mu^2}{3r_+^2}\right)\;.
\end{equation}
We can solve \eqref{bhtemp} for $r_+$ and substitute the result into $\eqref{fbh}$ to obtain $F_{bh}$ as a function of temperature and chemical potential.  With the scalar hair turned on, the free energy of the black hole is given by
\begin{equation}\label{fbhsc}
F_{bhsc}=\epsilon_{bhsc}-Ts-\mu\rho_{bhsc}\;,
\end{equation}
where the temperature and entropy can be obtained from \eqref{bhsctemp} and \eqref{bhscs}.  Since the critical temperature for the black hole to develop scalar hair is proportional to $\mu$,  this phase boundary is a line that passes through the point $(\mu,T)=(0,0)$.

Now we can begin to assemble our $T$-$\mu$ phase diagram.  The lowest value of $F$ among our four phases is favored. We assemble the phase diagram by the following procedure.  First, we analyze our system at a fixed value of $\mu$.  We must use \eqref{bhscaling} to scale our black holes with charged scalar so that all the solutions have the same value of $\mu$.   Having fixed all our solutions to have the same chemical potential, we now find the temperatures at which two of our solutions have the same free energy.  A phase boundary appears there if no other solutions have a lower free energy at that temperature.  We then repeat this with various values of $\mu$.  The results are shown in Figure \ref{phasediagrams}.  

\begin{figure}[h!]
  \centering
  \includegraphics[width=3in]{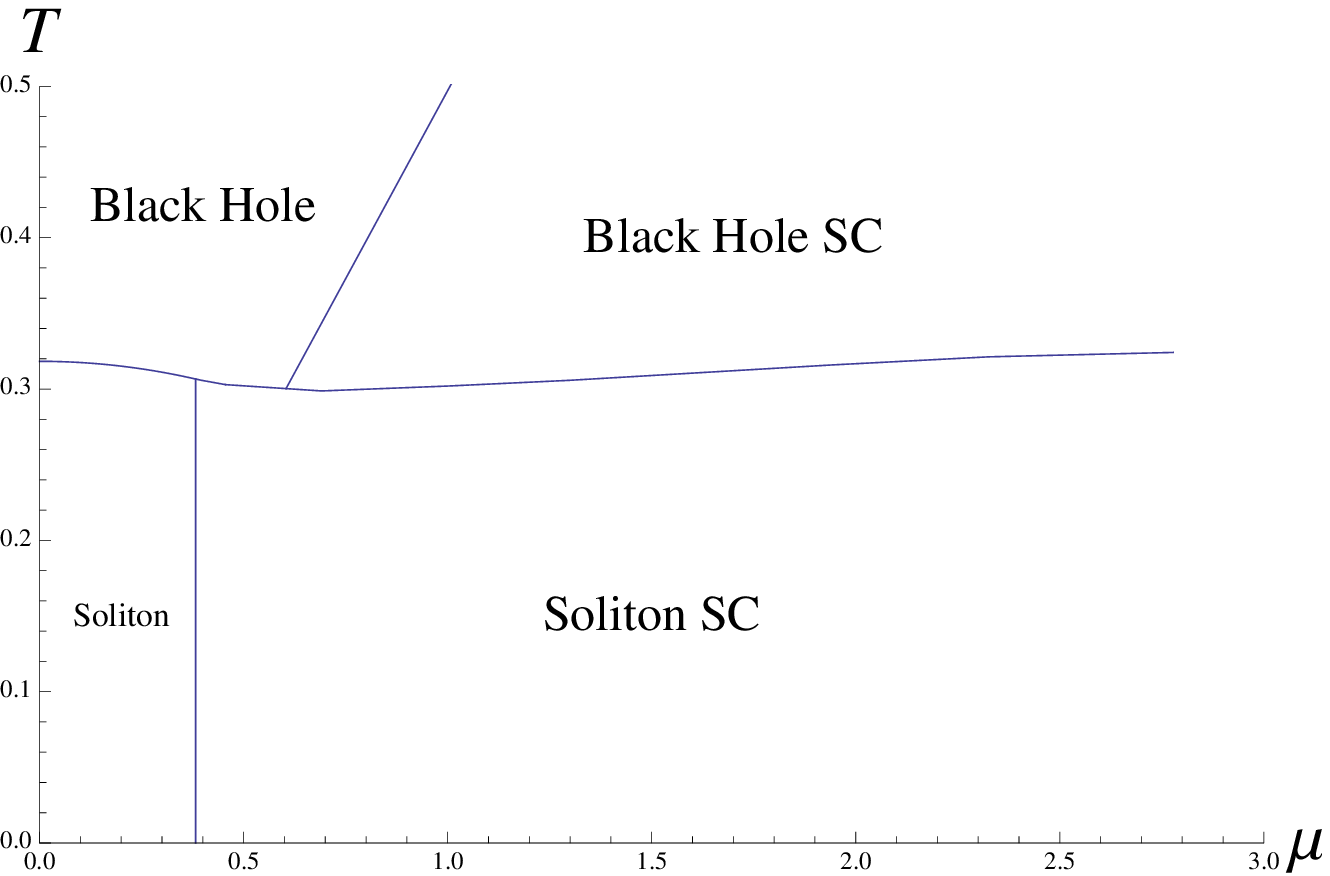}\qquad
  \includegraphics[width=3in]{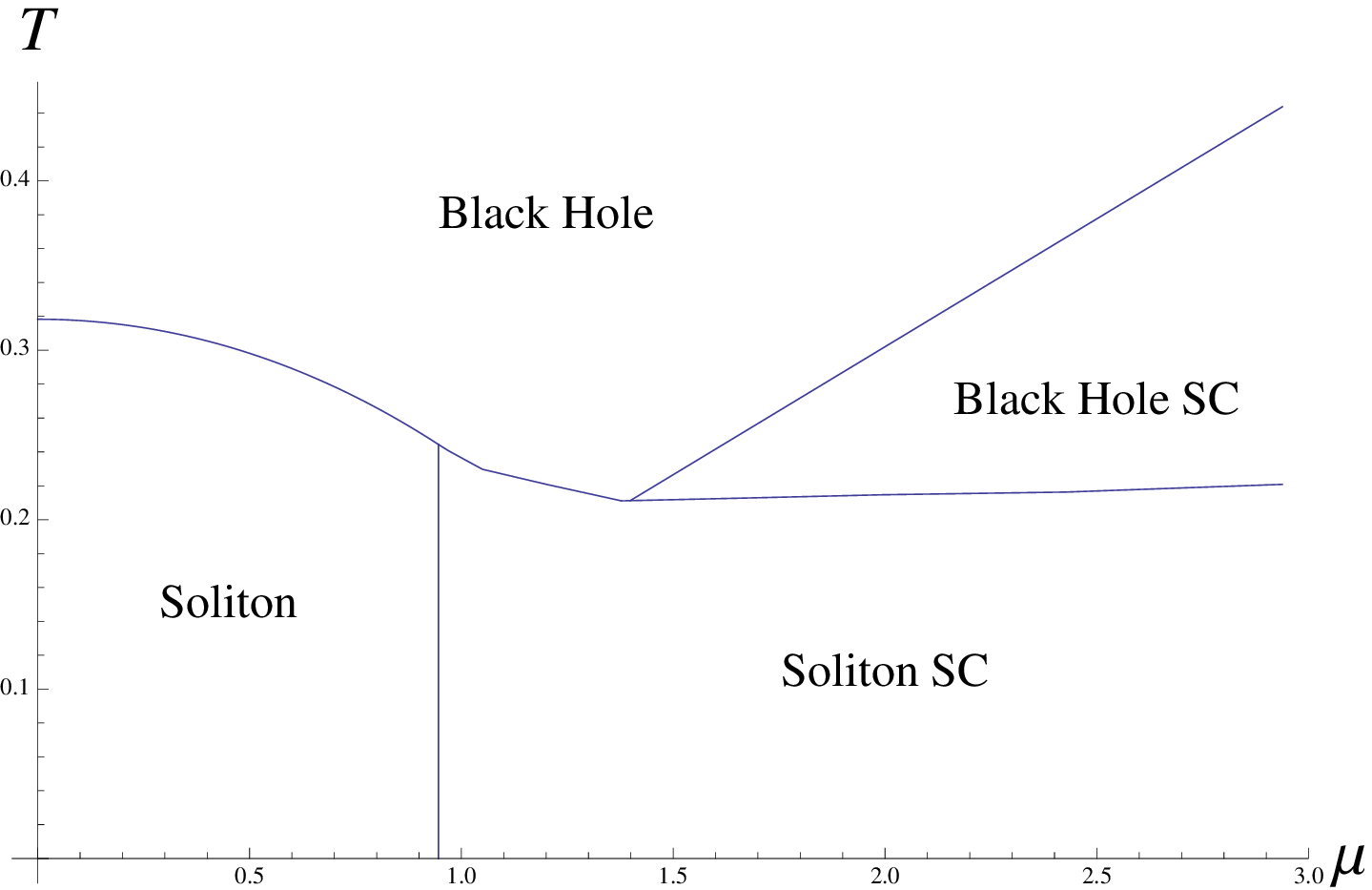}\qquad
  \linebreak
  \linebreak
  \includegraphics[width=3in]{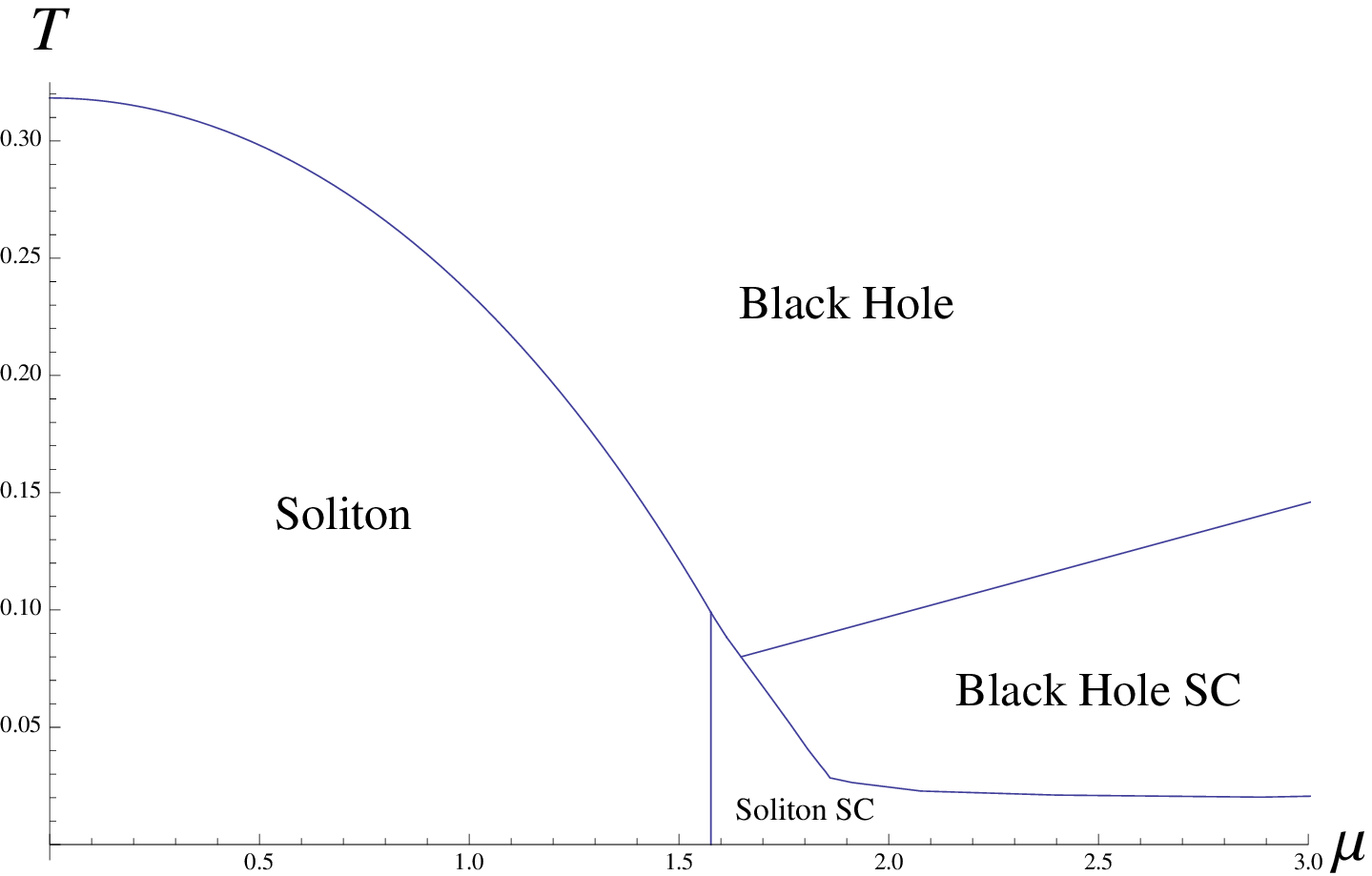}\qquad
    \includegraphics[width=3in]{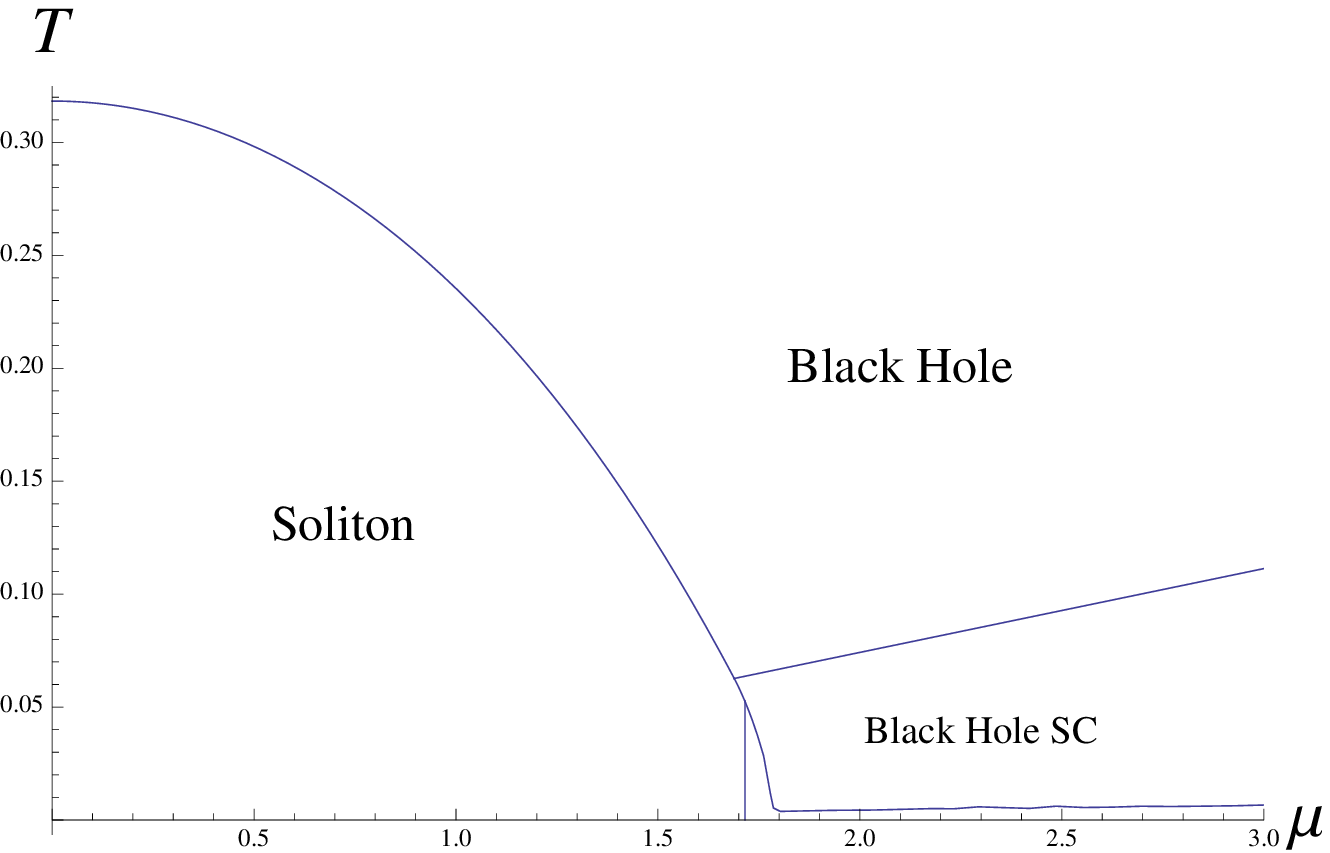}\qquad
    \linebreak
    \linebreak
  \includegraphics[width=3in]{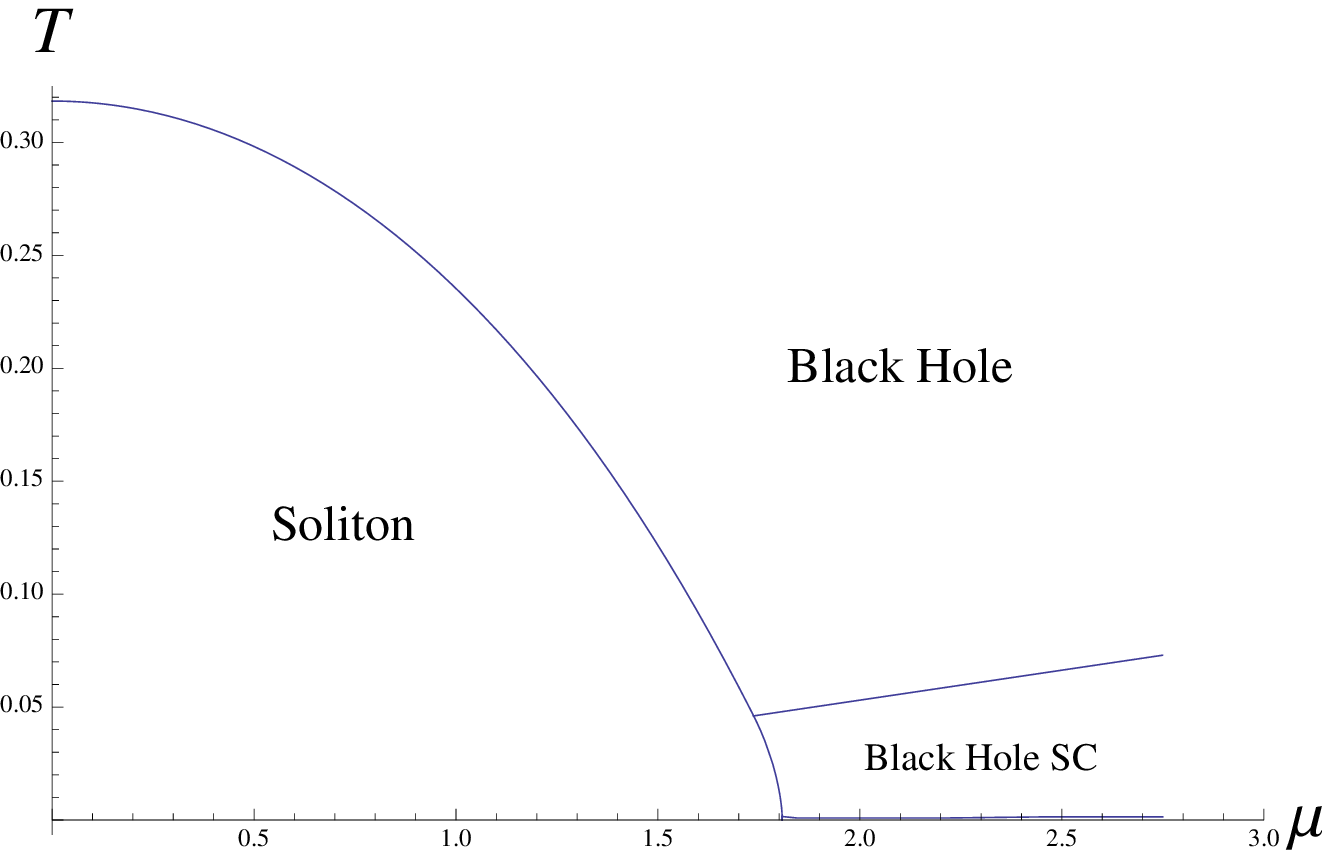}
  \caption{Starting from the top left, the phase diagrams for $q=5$, $q=2$, $q=1.2$, $q=1.1$, and $q=1$.  In the $q=1$ phase diagram, there is a phase boundary between the black hole superconductor and the soliton superconductor at about $T\sim.001$.  Here, we have scaled the period of the $\eta$ coordinate to be $\gamma=\pi$.  Note that for $q=1.1$ and $q=1$, there is a range of chemical potentials ($1.69\lesssim\mu\lesssim1.71$ for $q=1.1$ and $1.74\lesssim\mu\lesssim1.81$ for $q=1$) where we find a new  superconductor/insulator phase boundary.}
  \label{phasediagrams}
\end{figure}

Note that the phase boundary between the non-superconducting phases of the AdS black hole and the AdS soliton is known analytically.  At zero chemical potential, the boundary always occurs at $T=1/\pi$.    Also, if there were no superconducting phases, the phase boundary between the AdS black hole and the AdS soliton at zero temperature would be at a chemical potential $\mu=2^{1/2}3^{1/4}\approx 1.86$, in our units.


Unlike the analysis in the probe limit, the superconducting soliton phase seems to exist at all chemical potentials.  This is different from the case without scalars.  A possible physical explanation for this may be due to the fact that the curvature of these superconducting black hole becomes infinite at zero temperature (analogous to the four dimensional case \cite{Horowitz:2009ij}) and the system is trying to avoid singular configurations. We will discuss this further in the next section.

It is clear from Figure \ref{phasediagrams} that as we lower $q$, the phase boundaries behave as follows: (1) The line dividing the soliton insulator from the soliton superconductor moves to larger $\mu$. This is expected since a smaller charge requires a larger chemical potential to cause the scalar field to condense. (2) Similarly, the critical temperature for the black hole to develop scalar hair is lower as we reduce $q$. Thus, the slope of the  $T/\mu$ line marking this phase boundary is reduced. (3) The boundary between the black hole superconductor and soliton superconductor moves to lower temperature.

Interestingly, we find a peculiar phase structure for sufficiently low $q$ (e.g. $q\lesssim1.1$).  As one lowers $q$, the two triple points eventually merge into a quadruple point\footnote{At the quadruple point, the phase boundary between the two soliton phases is second order, which can be seen by comparing Figure 2 with Figure 6.} and then separate again.  Once they pass through each other, there is a small range of chemical potentials where a black hole conductor can be cooled to form a black hole superconductor and then cooled further to become a soliton insulator.  That is, there is a new first order superconductor/insulator phase transition that opens up at low temperature.  The free energy as a function of temperature is plotted in Figure \ref{fq1} at a fixed value of $\mu$ in this window.  

\begin{figure}[h]
  \centering
  \includegraphics[scale=1]{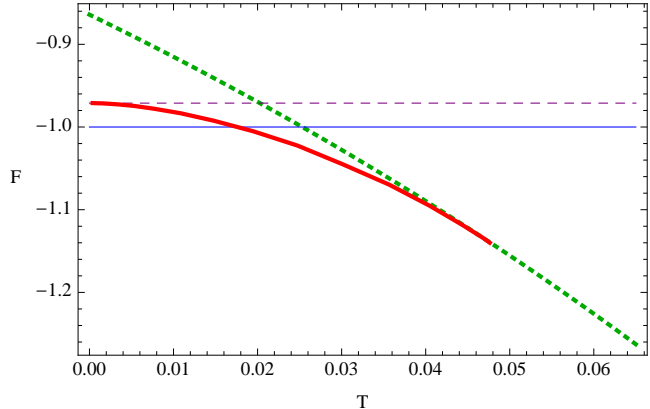}\qquad
  \caption{The free energy density vs temperature for the $q=1$ system at fixed $\mu\approx1.79$.  The blue horizontal line at $F=-1$ is the soliton insulator.  The dashed purple line at $F\approx-0.97$ is the soliton superconductor.  The dotted green line and the thick red line are the black hole and black hole superconductor, respectively.  Note that as we cool the system from high temperature, the free energy favors the black hole, then the superconducting black hole, then the soliton insulator.}
  \label{fq1}
\end{figure}

\end{section}

\begin{section}{Summary}

We have studied the four phases of the five dimensional  Einstein-Maxwell-scalar theory (\ref{action}) corresponding to black holes (with and without scalar hair) and solitons (with and without scalar fields). We found the solutions (with backreaction) for various charges $q $ for the scalar field. The black holes are always unstable to forming scalar hair along a line of constant $T/\mu$ exactly analogous to previous studies of the four dimensional case \cite{HHH08b}. The soliton is always unstable to turning on the scalar field along a line of constant $\mu$ as first shown in \cite{NRT09} for large $q$. These phase transitions are usually second order, however we found that the latter becomes first order for small $q$.

The complete phase diagrams were constructed for  $q \ge 1$ and shown in Figure 6.  In all these cases, the black hole is never the dominant configuration at low temperature.  As mentioned above, a physical reason for this could be that the  curvature of the hairy black hole becomes infinite at $T=0$, and the system is trying to avoid singular configurations.  One can test this by including a $\lambda|\psi|^4$ term in the action \eqref{action}.   With the added term, the hairy black hole will be nonsingular at $T=0$.  For small $\lambda$, the curvature is still large at low temperatures and the phase diagrams are expected be similar to those in Figure 6.  But for larger values of $\lambda$, the curvature is smaller at low temperatures and it is not clear whether or not the hairy black holes will be preferred at $T=0$.  

Although high curvature may be the reason the system with $q\ge 1$ prefers the soliton  over the black hole  at low temperature, it seems likely that this will change for $q < 1$. As we discussed in the previous section, the phase boundary, $\mu = \mu_c$, between the soliton and soliton superconductor increases as $q$ is lowered. But once $\mu_c > 1.86$, the extremal Reissner-Nordstrom AdS black hole has lower free energy  than the soliton. Since the hairy black hole will have even lower free energy, it seems clear that this solution must dominate the phase diagram for $T=0$ and $1.86 <\mu < \mu_c$.  In this case, the system has to approach the singular configuration since no other phase exists\footnote{An example of a zero temperature phase transition between a soliton and black hole is given in \cite{Imeroni:2009cs}.}. Higher numerical precision is required to compute the $q<1$ phase diagrams explicitly.  

Perhaps the most surprising feature of the phase diagrams is the existence of a new phase boundary between the hairy black hole and AdS soliton which opens up for $q\lesssim1.15$. This corresponds to a first order phase transition in which a superconductor turns into an insulator as the temperature is lowered. It would be interesting to know if any real materials behave this way. 

Finally, by considering the case of a neutral scalar field, we constructed the gravitational dual of a Bose Einstein condensate of glueballs, and found the maximum energy that can be supported this way without thermalizing.

\end{section}

\vskip 1cm
\centerline{\bf Acknowledgements}
\vskip .5 cm
It is a pleasure to thank T. Faulkner, S. Hartnoll, V. Hubeny,  R. Myers, M. Rangamani, and M. Roberts for discussions. This work was supported in part by the US National Science Foundation under Grant No.~PHY08-55415.

\singlespacing

\end{document}